\documentclass[preprint,
nofootinbib,
amsmath,
amssymb,
preprintnumber]{revtex4}

\usepackage{graphicx}
\usepackage{dcolumn}
\usepackage{bm}
\usepackage[dvipsnames]{xcolor}
\usepackage[utf8]{inputenc}
\usepackage{dcolumn}
\usepackage{color}
\usepackage{subfigure}
\usepackage{amsmath}
\usepackage{adjustbox}
\usepackage{float}
\usepackage{amssymb}

\begin{document}

\preprint{CYCU-HEP-24-02}

\title{A study of layered holographic superconductor}

\author{Chi-Hsien Tai}\thanks{xkp92214@gmail.com}
\affiliation{Department of Physics and Center for Theoretical Physics,\\
 Chung Yuan Christian University, Taoyuan, Taiwan}

\author{Wen-Yu Wen}\thanks{wenw@cycu.edu.tw}
\affiliation{Department of Physics and Center for Theoretical Physics,\\
 Chung Yuan Christian University, Taoyuan, Taiwan}
\affiliation{Leung Center for Cosmology and Particle Astrophysics, National Taiwan University, Taipei, Taiwan}

\begin{abstract}
We have investigated a holographic model of a multi-layered superconductor in (2+1)-dimensions using the AdS/CFT correspondence. This correspondence allows us to study strongly interacting condensed matter systems through a weakly interacting gravitational theory. Our study focused on the effects of a finite system size on the superconductor's properties.  We observed significant variations in the critical temperature and critical magnetic field depending on the number of layers and the spacing between them. Notably, our results capture the transition from a two-dimensional (2D) to a three-dimensional (3D) behavior in the limit of a large number of closely spaced layers. This transition signifies a change in how the superconductivity operates within real material.
\end{abstract}

\maketitle

\section{Introdution}
The holographic correspondence, initially proposed in the framework of AdS/CFT correspondence\cite{Maldacena:1997re, Gubser:1998bc, Witten:1998qj}, has expanded beyond gravitational theories and quantum field theories into various fields, including nuclear physics and condensed matter phenomena\cite{Herzog:2007ij, Hartnoll:2007ih, Hartnoll:2007ip, Hartnoll:2008hs, Minic:2008an}.  One significant application of this correspondence is in mapping superconductors to gravity duals \cite{Gubser:2005ih, Gubser:2008px, Hartnoll:2008vx}. This process involves introducing temperature by incorporating a black hole and a condensate through a charged scalar field. The key requirement is to have a system that allows for black holes with scalar hair at low temperatures while having no hair at high temperatures to accurately replicate the superconductor phase diagram.

{Expanding on this research, our paper delves into the finite size effects within a multi-layered superconductor. To model a multi-layered superconductor at finite temperature, we introduce a thermal background that corresponds to the near horizon limit of multi-centered membranes\cite{Gravity and Strings}.  Through our investigation, we have observed an intriguing phenomenon: the critical temperature increases as the number of layers in the superconductor grows. This observation suggests a potential crossover from $2$-dimensional to $3$-dimensional superconductivity as the system becomes more layered \cite{3D2DCrossOver, 3D2DCrossOver_finite size effect}.

By exploring these finite size effects in multi-layered superconductors, we aim to deepen our understanding of the underlying physics and potentially uncover new insights into the behavior of superconducting materials. This research contributes to the ongoing efforts to bridge the gap between gravitational theories and quantum field theories in various physical systems.

Our paper is organized as follows: in the section II, we utilize a membrane solution within eleven-dimensional supergravity to represent the layered system. Subsequently, we derive the equations of motion that govern the behavior of the relevant fields within this model.  In Section III, we delve into the effects of a finite system size on the superconductor's properties. This involves analyzing various parameters within our holographic model and observing their influence. Section IV then presents a comprehensive discussion of our findings.  We compare our results to both existing theoretical models and relevant experimental data to highlight the strengths and potential limitations of our approach.

\section{The multi-layered model}

Many unconventional superconductors have a layered crystal structure, like cuprates \cite{bednorz1986possible} and ion-based materials \cite{kamihara2008iron}. These materials exhibit unique properties due to their ($2 + 1$) dimensional nature. Beyond the basic behavior, we are now interested in how these layered structures interact with each other and how an applied magnetic field affects them.  To study these phenomena, we employ the powerful holographic tool of duality, which allows us to investigate a strongly-coupled system by studying a gravitational dual system in one higher dimension.  In this specific case, we start with a gravity model in ($3 + 1$) dimensions that is holographically equivalent to the desired layered superconductor in (2 + 1) dimensions. This superconductor exhibits superconductivity below a critical temperature \cite{Hartnoll:2008vx} and one was able to apply external magnetic field to observe the Meissner effect \cite{Nakano:2008xc, Albash:2008eh} and vortices \cite{Albash:2009iq}.

Here, we utilize a black M$2$-brane solution in $11$ dimensional supergravity, as the low energy limit of M-theory. This solution incorporates the multi-layered structure by including $y$ stacks of coincident M$2$-membranes, representing the individual layers \cite{Gravity and Strings}:
\begin{equation}\label{eqn:11dSUGRA}
ds^{2}=H_{M_{2}}^{-\frac{2}{3}}(-Wdt ^{2}+d\vec{x}^{2}+d\vec{y}^{2})+H_{M_{2}}^{\frac{1}{3}} (W^{-1}dr^{2}+r^{2}d\Omega_{(7)}^{2})
\end{equation}
\[W=1-\frac{\omega}{r^{6}}+\frac{M^{2}}{r^{8}},\ \omega=h(\alpha^{2}-1),\ h=\frac{(l_{Plank}^{(11)})^{6}}{6\omega_{(7)}}
\]
\[H_{M_{2}}=1+\frac{\frac{n}{y}h}{(r -d)^{6}}+\frac{\frac{n}{y}h}{(r -2d)^{6}}+...\frac{\frac{n}{y}h}{(r -(x-1)d)^{6}}
\]

We identify $n$ as the total number of branes which be equally distributed in $y$ stacks and $d$ as the thickness of insulating material in between layers. $\omega$ and $M$ represents the mass and magnetic charge of the black hole. In our numerical simulations, the size and temperature of the black hole are determined by adjusting $\alpha$, and $M$ plays the role of external magnetic field $B_{\perp}$ perpendicular to the brane.  In the following, we set the Planck length in 11 dimensions $l_{Plank}^{(11)}=1$, and the associated constant $h$ in the harmonic function can be regarded as the mass of a single brane in the stack.

At zero temperature, the BPS configuration of multi-layered solution can be obtained from separating the stack of coincident membranes without costing additional work, attributed to the gravitational attraction balances off the Ramond-Ramond force.
\\
\begin{figure}[H]
\begin{center}
\includegraphics[scale=0.7]{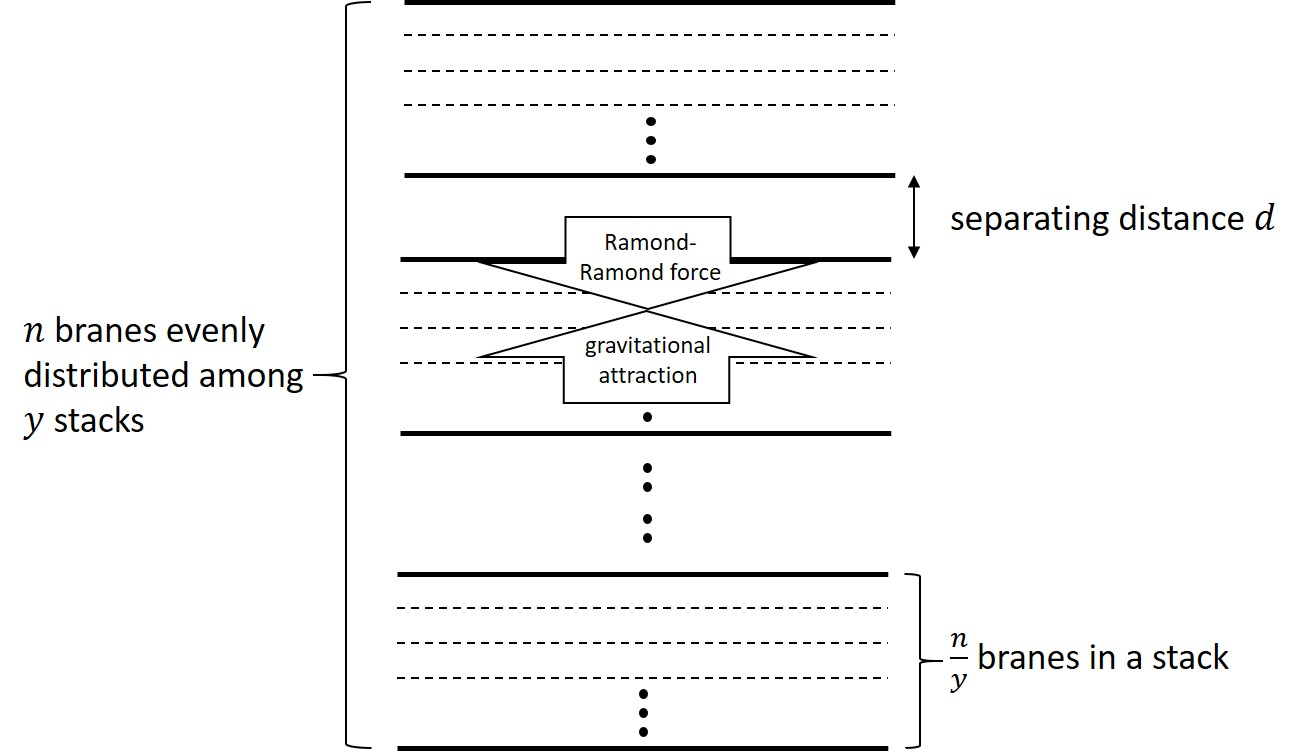}
\end{center}
\caption{\label{fig:multi-layered model}
{Schematic sketch of mulit-layered model.} }
\end{figure}

In the dual field theory side, this corresponds to the Coulomb phase of 2+1 super Yang-Mills theory by breaking gauge group $SU(n)$ into $SU(\frac{n}{y})\times	\cdot	\cdot	\cdot\times SU(\frac{n}{y})$. Nevertheless, this balance is broken due to a black hole at the center for heating up the system and adding external magnetic field.  Additional components of $C_{mnp}$ is needed to sustain this multi-layered configuration at finite temperature. We used this layer structure to examine how those variables affect the superconductor's critical temperature and to further investigate their possible connection to the real world materials.

We define the distance between layers $d$ to be much smaller than $r$ and apply Taylor expansion to obtain a more concise form of the harmonic function:
\begin{equation}\label{eqn:H}
H_{M_{2}} \simeq H = 1+\frac{n h}{r^{6}}+\frac{3(y+1)\frac{n}{y}h}{r^{7}}d+O(d^{2})
\end{equation}

When the layers are all coincident, say $d=0$, in the near-horizon limit $r \ll 1$, by using the transformation $r=(n h)^\frac{1}{12}\rho^{\frac{1}{2}}$, one reproduces (\ref{eqn:11dSUGRA}) as a product space of the seven sphere and magnetically charged black hole in $AdS_{4}$, which reads:

\begin{equation}\label{eqn:AdS4}
ds^{2}=-f(\rho)dt^{2}+\frac{d\rho^{2}}{f(\rho)}+\frac{\rho^{2}}{h^{\frac{1}{3}}}(dx^{2}+dy^{2}),    
\end{equation}

where 
\begin{equation}f(\rho)=\frac{\rho^{2}}{n^{\frac{1}{3}}h^{\frac{2}{3}}}(1-\frac{\omega}{n^{\frac{1}{2}}h^{\frac{1}{2}}\rho^{3}}+\frac{M^{2}}{n^{\frac{2}{3}}h^{\frac{2}{3}}\rho^{4}}).
\end{equation}

The radius of curvature of $AdS_{4}$ space is $L= n^{\frac{1}{6}}h^{\frac{1}{3}}$.

In the following, we well further rescale the radial direction by using the transformation $\rho=(\frac{n}{h})^{\frac{1}{6}}z^{-1}$ , which converts infinite distance to a finite value to facilitate numerical computation.

The AdS metric in (\ref{eqn:AdS4}) is slightly deformed by small but nonzero $d$.  In previous applications of AdS/CFT correspondence, conformal symmetry was broken while maintaining its validity. For instance, in the holographic model of QCD, confinement was achieved by introducing a scalar field (soft-wall) or a sharp cutoff (hard-wall)\cite{Andreas:2006, Leandro:2006, Erlich:2005}. Here, we adopt an open-minded approach and assume the AdS/CFT correspondence remains valid under a deformation parameterized by $d$.  By solving the equation $W(z_{H})=0$, we can determine the horizon position $z_{H}$. The temperature of multi-layer system is again given by Hawking temperature of black hole:
\begin{equation}
T_H=\frac{\sqrt{\partial_{z}F(z_{H})\times\partial_{z}G(z_{H})}}{4\pi},
\end{equation}
where 
\begin{eqnarray}
F(z_{H})&=&W(z_{H}) H(z_{H})^{-\frac{2}{3}},\nonumber\\
G(z_{H})&=&W(z_{H}) H(z_{H})^{\frac{1}{3}}.
\end{eqnarray}

Now we are ready to probe this backrgound with the Ginzburg-Landau (GL) action for a Maxwell field and a charged complex scalar, which does not back react on the metric. The Lagrangian density is given by
\begin{equation}\label{eqn:probed_action}
\mathcal{L}_{m}\sim-\frac{1}{4}\mathcal{F}_{ab}\mathcal{F}^{ab}-\frac{2}{L^{2}}\Psi^{*}\Psi-g^{\mu\nu}(\partial_{\mu}\Psi^{*}-iA_{\mu}\Psi^{*})(\partial_{\nu}\Psi-iA_{nu}\Psi).
\end{equation}

We assume the planar symmetry ansatz for all fields: the complex scalar $\Psi=\Psi(z)$, the scalar potential $A_{t}=\Phi(z)$ and components of vector potential $A_{x}=B_{\parallel} z$, $A_{z}=A_{y}=0$, where $B_{\parallel}$ represents the parallel magnetic field on the brane. Then we need to solve the equations of motion for $\Psi$ and $A_t$:

\[
\begin{aligned}
\frac{\partial L}{\partial \Psi^{*}} =0&= -\sqrt{-g}\frac{2}{L^{2}}\Psi+\sqrt{-g}(\frac{-1}{\sqrt{-g}})\partial_{\mu}[\sqrt{-g}g^{\mu\nu}(\partial_{\nu}\Psi)-iA_{\nu}\Psi]-i\sqrt{-g}A_{\mu}g^{\mu\nu}(\partial_{\nu}\Psi-iA_{\nu}\Psi)  &\\ 
& =\frac{-1}{\sqrt{-g}}\partial_{\mu}[\sqrt{-g}g^{\mu\nu}(\partial{\nu}\Psi-iA_{\nu}\Psi)]+iA_{\nu}g^{\mu\nu}(\partial_{\nu}\Psi-iA_{\nu}\Psi)-\frac{2\Psi}{L^{2}},
\\
\frac{\partial L}{\partial A^{*}}=0 &=-\frac{1}{\sqrt{-g}}\partial_{z}(\sqrt{-g}\partial_{z}A_{t})-2 g^{tt}A_{t}\Psi^{2}. \\
\end{aligned}
\]

For simplicity we have fixed the phase of complex scalar field to be zero. Keeping the real part, we can get a pair of coupled second order differential equations:
\begin{equation}
\begin{aligned}
&\Psi''+(\frac{W'}{W}-\frac{2}{z})\Psi'+\frac{h^{\frac{1}{3}}}{2z^{3}}\frac{H^{\frac{1}{3}}}{W}\Psi+\frac{h^{\frac{1}{3}}}{4z^{3}}\frac{H}{W^{2}}\Phi^{2}\Psi+\frac{h^{\frac{1}{3}}}{4z}\frac{H}{W}B_{y}^{2}\Psi=0, \\
&\Phi''+(\frac{2H'}{3H}-\frac{2}{z})\Phi'+\frac{h^{\frac{1}{3}}H^{\frac{1}{3}}}{2z^{3}W}\Psi^{2}\Phi-\frac{h^{\frac{1}{3}}H^{\frac{1}{3}}}{2z^{2}W}B_{y}\Psi^{2}\Phi+(\frac{2}{z}-\frac{2H'}{3H})B_{y}W-W'B_{y}=0.
\end{aligned}
\end{equation}

We recall that the harmonic function $H$, given by (\ref{eqn:H}), is the $H_{M_{2}}$ in the near-horizon limit and retain only the first-order term of $d$ in the expansion.  As a consequence, our simulation results will be influenced by the separation distance $d$ and number of layers $y$.

In particular, for normalizable scalar perturbation, we require at the horizon:
\begin{equation}
\begin{aligned}
&\Phi(z_{H})=0, \\
&\Psi(z_{H})=-\frac{3}{2} \Psi'(z_{H}).
\end{aligned}
\end{equation}
At the boundary, they behave like
\begin{equation}\label{eqn:Psi,Phi Ansatz}
\begin{aligned}
&\Psi=\Psi_{(1)}z+\Psi_{(2)}z^{2}+...,\\
&\Phi=\mu-\rho z+...,
\end{aligned}
\end{equation}
where $\mu$ and $\rho$ are interpreted as chemical potential and charge density in the dual field theory. We are interested in the case where either $\Psi_{(1)}$ or $\Psi_{(2)}$ vanishes for stability concern at the asymptotic region of AdS, then the nonzero $\Psi$ gives the condensation:
\begin{equation}
\begin{aligned}
\langle O_{i} \rangle=\sqrt{2} \Psi_{(i)},  i=1,2
\end{aligned}
\end{equation}

\section{Parameters for multi-layer superconductor}

In the normal phase, we always have solutions to the equations (\ref{eqn:Psi,Phi Ansatz}), that is $\Psi=0$ and $\Phi=\mu-\rho z$; while in the superconducting phase, we may have nontrivial $\Psi(z)$ and its boundary value serves as an order parameter for condensate. In the absence of applied magnetic field, for any fixed $\rho$, there exists a critical temperature $T_{c}$, above which there is no more nontrivial solution. 

\begin{figure}[H]
\begin{center}
\includegraphics[scale=0.9]{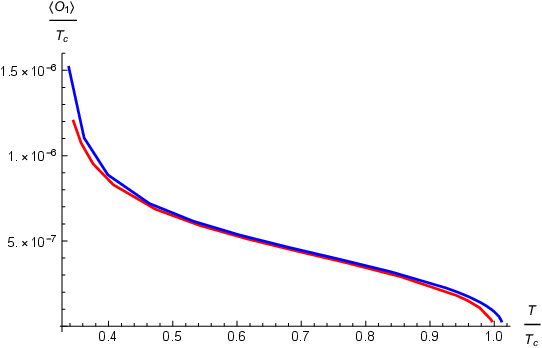}
\includegraphics[scale=0.8]{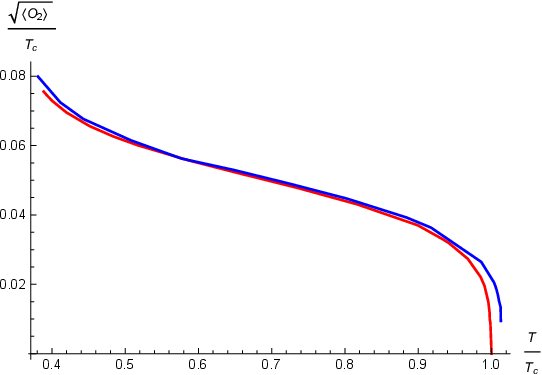}
\end{center}
\caption{\label{fig:n100TcPsi1}
Condensation as a function of temperature for two operators $O_{1}$ (left) and $O_{2}$ (right) in layered structure. The red curve shows single layer ($d=0$) and blue curve indicates multi-layered structure ($d=0.01$). Those curves are normalized by the critical temperature at $d=0$. The rising of critical temperature in the layered superconductor implies a $2$d to $3$d transition.} 
\end{figure}

In the Figure \ref {fig:n100TcPsi1}, we compare the condensation as a function of temperature for both single and multi-layered superconductors. It shows that layered structure plays a role of {\sl thickness} to make the planar superconductor $3$D-like.  As a consequence, the critical temperature rises up for multi-layered structure.\\

\subsection{Charge density}

The metric (\ref{eqn:11dSUGRA}) describes a total number of branes, denoted as $n$, evenly distributed among $y$ stacks.  The brane number $n$ is positively correlated with charge density in the $2$D superconductor, as shown in the Figure \ref{fig:branenumber}.  In the superconducting phase, the density of Cooper pairs should be proportional to the charge density.  Therefore we expect that the critical temperature rises with brane number as well, which is also confirmed in Figure \ref{fig:branenumber}.  In practice, parameter $n$ could be tuned for different superconducting material with specific charge density.

\begin{figure}[H]
\begin{center}
\includegraphics[scale=0.8]{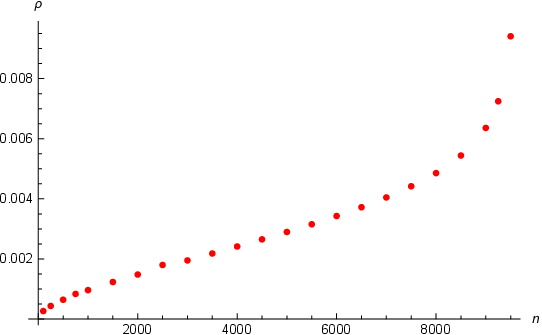}
\includegraphics[scale=0.8]{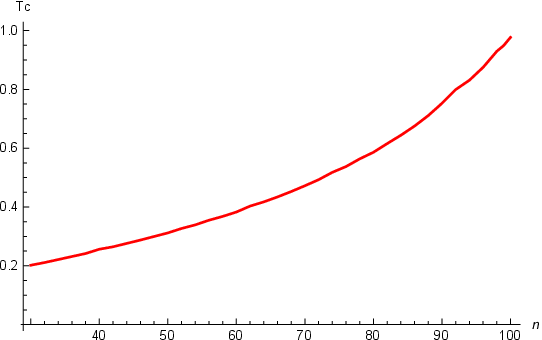}
\end{center}
\caption{\label{fig:branenumber}
The charge density (left) and critical temperature (right) are positively correlated with total number of branes $n$.}
\end{figure}

\subsection{Inter-layer distance}

\begin{figure}[H]
\begin{center}
\includegraphics[scale=0.8]{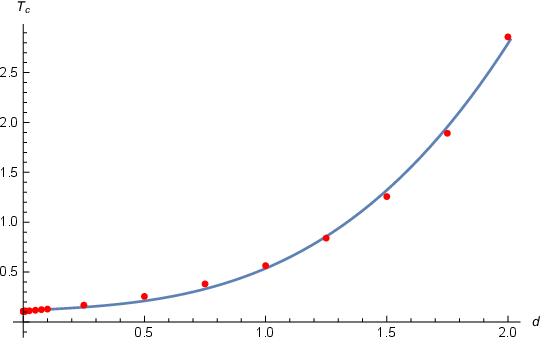}
\includegraphics[scale=0.8]{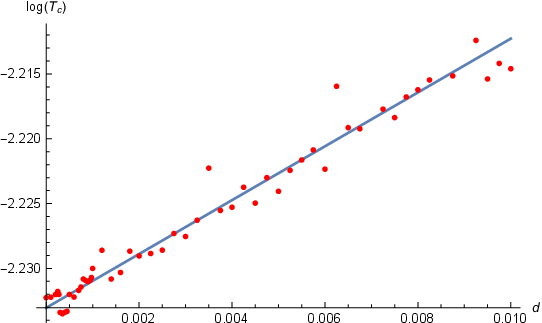}
\end{center}
\caption{\label{fig:distance}
Both figures depict the critical temperature grows exponentially as a function of inter-layer distance $d$ at large (left) and small scale (right).  In particular, $\Delta T_c \propto e^{2.07d}$ in the small $d$ regime.}
\end{figure}

The inter-layer distance $d$ corresponds to the thickness of the insulating layer in multi-layer superconductors.  In the Figure (\ref{fig:distance}), the critical temperature appears to increase exponentially with the inter-layer distance.  Although we are reminded that only the regime with small deformation $d \ll r$ can be trusted according to (\ref{eqn:H}). We nevertheless observe that $T_c(d) \simeq T_c(0) e^{2.07d}$ in that regime.

\subsection{Critical magnetic field}
The Meissner effect is expected for the holographic superconductor in the presence of an applied magnetic field \cite{Nakano:2008xc,Albash:2008eh}.  That is, there exist both $T_{c}$ and a critical magnetic field $M_{c}$, above that the nontrivial solution is not admissible.

\begin{figure}[H]
\begin{center}
\includegraphics[scale=0.9]{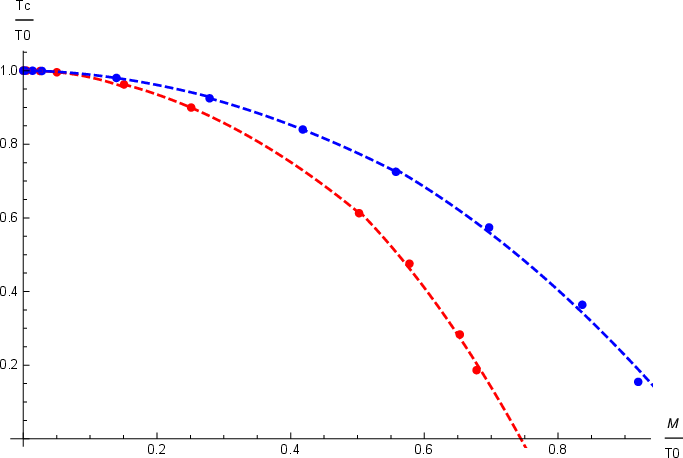}
\end{center}
\caption{\label{fig:Phase diagram}
Phase diagram of $T-M (B_\perp)$ for two different numbers of layers, where superconducting (normal) phase exists in the region below (above) the fitting curves.  Blue dots (stack number $y=100$) and red dots ($y=10$) represent different number of layers.  It shows that given the same critical temperature, it has better resistance to an external magnetic field as number of layers increases.}
\end{figure}

In the Figure \ref{fig:Phase diagram} we show phase diagram for two different numbers of layers.  It shows that given the same critical temperature, it has better resistance to an external magnetic field as number of layers increases, given the same inter-layer distance.  In practice, this suggests an enhancement of critical magnetic field for increasing thickness, as expected crossover from $2$D to $3$D.

\color{black}

\begin{figure}[H]
\begin{center}
\includegraphics[scale=0.9]{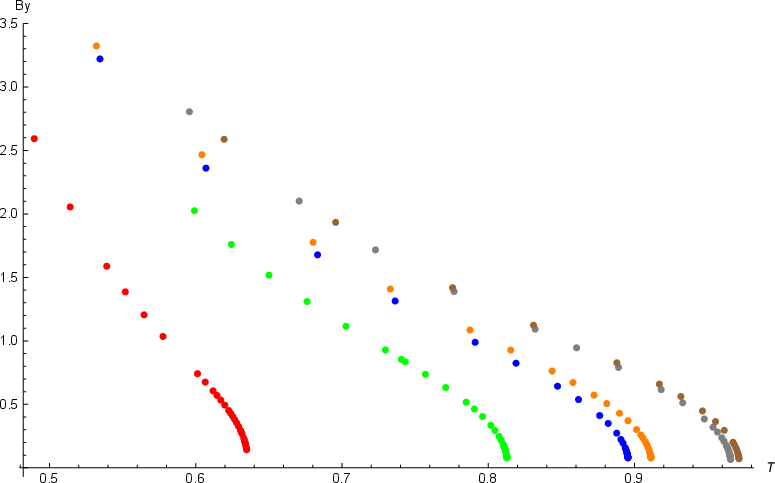}
\end{center}
\caption{\label{fig:Applied magnetic field}
The graph describes the relationship between temperature ($T$) and parallel magnetic field ($B_y$) for various stacking configurations given the same maximum thickness and brane number ($n=10^4$). From left to right, stack number $y=2$ (red), $4$ (green), $8$ (blue), $10$ (orange), $100$ (gray), $10^4$ (brown).}
\end{figure}

As shown in the Figure \ref{fig:Applied magnetic field}, we distribute the same number of total branes into different stacks.  We obaserve that critical temperature rises with numbers of stacks.  We regard the minimum stacks $(y=2)$ configuration as $2$D-like for large separation in between stacks, while the maximum stacks $(y=10^4)$ one as $3$D-like for denser packing of conducting layers.  We conclude that critical temperature is also enhanced from $2$D to $3$D.

\begin{figure}[H]
\begin{center}
\includegraphics[scale=0.8]{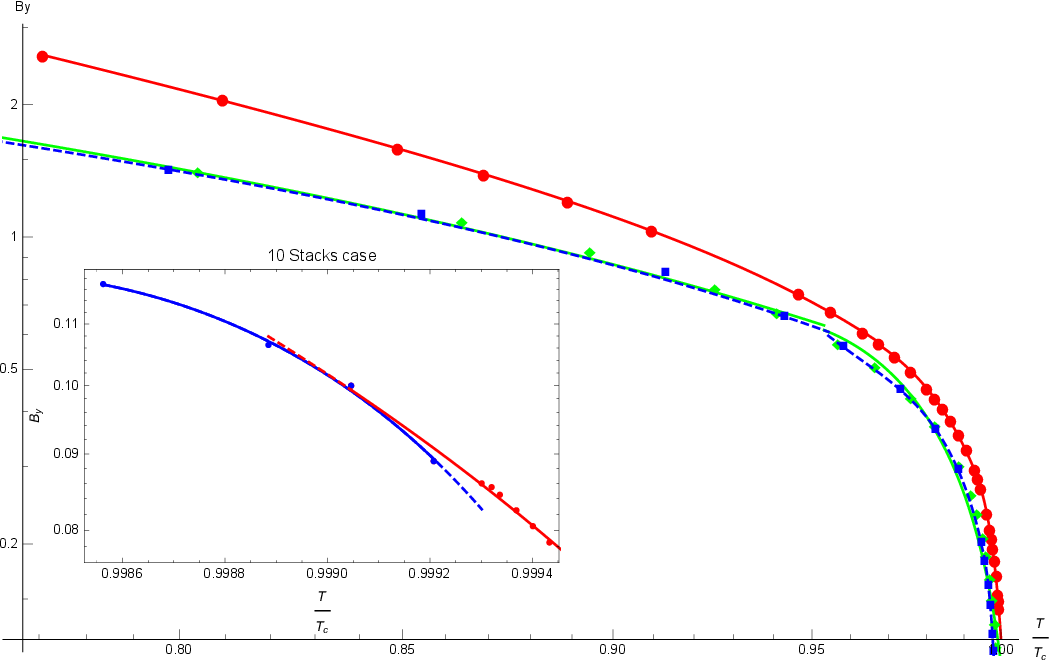}
\end{center}
\caption{\label{fig:2.5D}
Normalized plots of Figure (\ref{fig:Applied magnetic field}) for $y=2$ stacks($\bullet$), $y=10$ stacks($\blacklozenge$), $y=10^4$ stacks($\blacksquare$) to represent the crossover from $2$D to $3$D.  The inset shows the middle curve $y=10$ could be better fitted by two disconnected curves near the critical temperature. Similar dimensional crossover was observed in the Pb/Ge multilayer system \cite{Bruynseraede:1992}. }
\end{figure}

After normalizing the curves in the Figure (\ref{fig:Applied magnetic field}) with each critical temperature, we obtain the Figure (\ref{fig:2.5D}). We can manage to fit $B_{y} \propto (1-T/T_{c})^{\alpha}$ to fit all the data points, where $\alpha$ increases from $0.457$ to $0.566$ with $y$.  However, if we zoom in those points near the critical temperature, the value of $\alpha$ can have a wider range from $0.251$ to $0.986$. That the  exponent approaches one indicates the curves become more {\sl linear} and the multi-layer system behaves more like a $3$D geometry. It is tempting to refit one set of data points ($y=10$) by two disconnected curves; a square-root curve joins with a linear one.   We remark that similar dimensional crossover was observed in the Pb/Ge multilayer system\footnote{In the work done by Y. Bruynseraede et al. \cite{Bruynseraede:1992}, they observed the dimensional crossover in a set of Pb/Ge samples by varying Ge (insulator) thicknesses and fixed Pb thickness to $140$\r{A}.  They reported the upper critical magnetic field had the similar disconnected fit when Ge thickness was $20$\r{A} in the Figure ($2$) in their paper.}

\section{Discussion}
We have constructed a holographic model of a multi-layered superconductor existing in ($2+1$) dimensions. This model is based on a multi-membrane solution within the eleven-dimensional framework of M-theory. Using numerical simulations, we demonstrate a positive correlation between the density of charge carriers (Cooper pairs) and the number of branes in the model. This correlation directly impacts the critical temperature, the temperature at which the material transitions into a superconducting state. Furthermore, by varying the number of layers (stacks) and the separation distance between them, we observe a dimensional crossover, particularly evidenced by the behavior of the parallel critical magnetic field.  We have following remarks:
\begin{itemize}
    \item The analysis of the non-BPS solution at finite temperature (equation (\ref{eqn:11dSUGRA})) reveals the necessity of incorporating additional three-form components. These extra components might play a crucial role in mimicking the effect of stress on bridging the layers within the actual material.  Investigating the precise relationship between these components and the underlying lattice stress in the real superconductor presents a fascinating avenue for future research. Determining if these components directly encode the stress or influence it through a more complex interaction could provide valuable insights into the interplay between geometry and material properties in this system. 
    \item Our study highlights an interesting asymmetry in how we introduce perpendicular and parallel magnetic fields into the model. The perpendicular component is incorporated through the background metric (equation (\ref{eqn:11dSUGRA})), while the parallel component is introduced through a separate term in the action (equation (\ref{eqn:probed_action})). This difference in treatment might be unnecessary.  While our simulations of the parallel critical field at the dimensional crossover appear consistent with experimental results \cite{Bruynseraede:1992}, the perpendicular critical field remains unaffected, as predicted by Ginzburg-Landau mean field theory. Ideally, for consistency, both components should be introduced in the same way, either within the background or the probed action. Further investigation is needed to explore this and potentially unify the treatment of both magnetic field directions.
    \item In this study, we utilize a holographic model of an s-wave superconductor due to its relative simplicity. While this model provides valuable insights, it's important to acknowledge that the majority of high-temperature superconductors exhibit d-wave pairing symmetry, as evidenced by research such as \cite{Hirschfeld:1994}. To achieve a more comprehensive understanding, extending our work to a holographic model that captures the d-wave behavior (such as \cite{Chen:2010mk, Benini:2010qc}) would be highly desirable. 
    \item From the perspective of effective field theory, a Ginzburg-Landau approach can model multi-layered superconductors using two or more order parameters, each representing a distinct layer within the material \cite{EabTang:1988}. This highlights the presence of independent superconducting behavior in each layer. Interestingly, this concept of multiple order parameters resonates with holographic studies on multi-band superconductors such as \cite{Wen:2013ufa}. Investigating whether these holographic models can be extended to capture the specific context of multi-layered structures presents a compelling avenue for future research.
\end{itemize}

\begin{acknowledgments}
Some parts of this work were reported by C. H. Tai in the 2022 Annual Meeting of the Physical Society of Taiwan.  This work is supported in part by the Taiwan's National Science and Technology Council (NSTC 109-2112-M-033-005-MY3 and 112-2112-M-033-003-MY3) and the National Center for Theoretical Sciences (NCTS).
\end{acknowledgments}

\bibliography{apssamp}

\end{document}